\documentclass{article}
\usepackage{graphicx}
\usepackage{epstopdf}
\usepackage[all]{xy}

\newcommand{\qed}{\nobreak \ifvmode \relax \else
      \ifdim\lastskip<1.5em \hskip-\lastskip
      \hskip1.5em plus0em minus0.5em \fi \nobreak
      \vrule height0.75em width0.5em depth0.25em\fi}

\begin{document}

\title{On  pricing kernels, information and risk}
\author{Diane Wilcox\ddag\footnote{School of Computational \& Applied Mathematics,
University of the Witwatersrand, Johannesburg, Private Bag 3,
Wits, 2050,  South Africa, \dag {\tt tim.gebbie@physics.org} \ddag
{\tt diane.wilcox@wits.ac.za}} and Tim Gebbie\dag }

\maketitle

\begin{abstract}
We discuss the finding that cross-sectional characteristic based
models  have yielded portfolios with higher excess monthly returns
but lower risk than their arbitrage pricing theory  counterparts
 in an analysis of equity returns of stocks listed on the JSE.
Under the assumption of general no-arbitrage conditions, we argue
that evidence in favour of characteristic based pricing implies
that information is more likely assimilated  by means of nonlinear
pricing kernels for the markets considered.
\end{abstract}


%
\begin{center}{\footnotesize{Keywords: Arbitrage pricing theory, characteristic based models, size effect, value effect, linear pricing kernel, nonlinear pricing kernel}}
\end{center}


\section{Introduction}

In neo-classical finance, future securities prices are regarded as
a function of risk and (realisable) present  value. We  compare
{\em arbitrage pricing theory} (APT) risk-factor pricing models
with {\em characteristic based models} (CBM), which offer
alternative causal relationships for returns in a equity markets.

 {\em No-arbitrage} (NA) refers to the
reasonable idea that a (theoretic) riskless portfolio with
possible { positive return}  in the future, must cost something
now. It can be shown to be equivalent to the existence of a
positive linear pricing rule or {\em pricing kernel}
\cite{HK1979,Ross1976} and has been generalised to the concept of
\emph{no-free-lunch-with-vanishing-risk} \cite{DS1994}, in an
overarching valuation framework \cite{AG1994} that has facilitated
the growth of a multi-trillion dollar derivatives market concerned
with pricing future cashflows, risk-sharing and completing
markets.

The general linear factor APT model formulation was introduced by
Ross to explain returns in the simplest possible manner consistent
with the assumption of NA \cite{DR2003, Ross2005}. The
Fama-and-French model followed thereafter, as special case, to
incorporate specific risk premia which were not explained by the
CAPM, namely {\em value} and {\em size}. In particular, Ball, Banz
and Basu \cite{Ball1978,Banz1981, Basu1983}  were amongst the
earliest authors to discuss empirical evidence that
small-capitalised stocks and stocks with low price-to-book values
exhibit higher long-term performance. More than a decade of
literature demonstrating the explanatory power of stock specific
accounting variables followed before Haugen and Baker eventually
crystallised the approach in the literature by describing a
general characteristic based model for equity market returns
\cite{HB1996}. This presented a different paradigm to the
 Fama-and-French time-series factors, which had been constructed to be consistent with a linear regression approach for
explaining risk in both stocks and bonds and which could be combined
with macro-economic variables in the APT framework.

In a comprehensive triple-sort testing framework to separate overlapping effects,  Daniel and Titman
 compared characteristic based models and Fama-and-French APT models
with attention to the same market factor, size and value variables
\cite{DT1997}. They show that the return premia to size and value do not appear to
originate from the covariance of stocks with intrinsic risk factors,
but that cross-sectional price variation seems to be driven directly by
stock characteristics. In particular, they demonstrate
portfolios which have similar characteristics (size and
book-to-market), but significantly different loadings to the Fama-and-French
factors can have similar average returns.

One of the findings in \cite{DT1997}  is that $\alpha$ {\em is
generally non-zero } for both models, contradicting the FF model's
linear requirements. Similar results were obtained in
\cite{G2007}. A possible reconciliation of their findings is that
$\alpha$ and the factor $\beta's$ are nonlinear functions of risk.
Ferson and Harvey \cite{FH1999} used out-of-sample testing to
illustrate the effectiveness of time-varying alpha and beta
APT-type models.

We adopt the predictive form of the models used in \cite{FH1999} for a simplified comparison of {\em conditional} APT
models\footnote{ i.e. with time-varying alpha and beta's} with
corresponding CBM models\footnote{The authors of this report are
not aware of other investigations with this particular innovation
in the literature.} with respect to their abilities to forecast
returns. This avoids the cumbersome triple-sorting as in \cite{DT1997}.
In particular we compare the returns of portfolios whose
weights (loadings) are updated monthly according to out-of-sample
expected parameters and then select stocks with the best expected
returns predicted using (2) and (4) below.

We consider portfolios whose weights (loadings) are updated monthly according to out-of-sample
expected returns according to equations (2) and (4) below.
The comparable APT and CBM implementations are constructed
using stocks listed on the South African Johannesburg Stock
Exchange (JSE)  between 1994 and 2007. We
demonstrate that characteristic-based models have provided higher returns at lower
volatility than risk-based models over a relatively short time
horizons.

Since we restrict our attention to specific sets of information in our comparisons, the out-performance suggests that CBM are more efficient in pricing in those types of information.


Thus, the characteristic approach not only provides a
empirical tool for exposing the phenomenology and structure of a
given market,  is it is also useful for making ex-ante stock
return predictions \cite{HB1996}. This makes the characteristic
approach interesting both practically and theoretically.


We discuss our findings in the context of pricing risk under assumptions of {\em no-arbitrage} and the weaker {\em no-free-lunch-with-vanishing-risk}: In
Section 2 we review the two model types and evolution of ideas, as
motivated by empirical evidence for equity returns.  In Section 3
we review the market context for our investigation and in Section
4  we present the outcomes of our analysis of the risk-return
profiles for portfolios constructed via the different asset price
models. We conclude in Section 5.

\section{Models for equity returns and risk premia}

The idea that NA drives pricing is coupled to the perspective that
risk factors exogenous to the market can move prices and that news
arrives often and randomly and gets priced in instantaneously. In
particular, this theory assumes that prices incorporate the impact
of exogenous factors fully. If this is the case, then endogenous
factors should be computable from the reflection of exogenous
factors in price changes (see for example \cite{Matoti2009,
WG2008}). Notably, APT does not specify its risk factors and
macroeconomic, fundamental or statistical factors may be used,
provided the dependence on these is linear .

Endogenous characteristics (attributes), such as  book-to-price or earnings-per-share, may also be used to
construct portfolios that represent the risk factors used in a
linear pricing rule \cite{Ross1976}.
 Fama and French \cite{FF1992, FF1996} developed a
model to compensate for two such attributes which have been
correlated with returns, namely value (measured via
book-to-price) and size.

The initial idea of characteristic based models was that company
specific characteristics could directly explain most of the
expected return differentials. However, to facilitate consistent pricing of uncertain future cash-flows of different types of securities,  factor mimicking portfolios that
mirror the roles of  characteristics, were devised.  The use of stock characteristics to generate a covariance structure
was a key  innovation in asset pricing \cite{FF1993}, since it allowed
a parsimonious interpretation without the rejection of linear
pricing models and the resulting 3-factor FF risk-based
model became a standard against which others were compared


\subsection{APT risk factor vs.  characteristic based
models}

We describe the two classes of models which are compared in order
to identify how information and risk are reflected in equity
prices.

 For an APT-type
risk factor model, we implement the following to explain and
predict returns:

\begin{eqnarray}
R_{i,t} = \alpha_t^F +\sum_j\beta_{i,j,t} f_{j,t-1} + \epsilon_{i,t}.\label{eqn:risk}
\end{eqnarray}
Here the unexplained component $\epsilon_{i,t}$ relates to the
factor loading (coefficient/weight) of $i$-th stock with respect to
the return on $j$-th price risk factor $f_{j,t}$, at time $t$, and
the risk factors are Fama-and-French model risk factors, derived
from size and value information, which are described in the next
section. The expectation of return at time $t-1$ for time $t$ is:
\begin{eqnarray}
\mathrm{E}_{t-1}[R_{i,t}] = \mathrm{E}_{t-1}[\alpha_t^F] + \sum_k \mathrm{E}_{t-1}[\beta_{i,j,t}] f_{j,t-1} \label{eqn:exprisk}
\end{eqnarray}

 Haugen and Baker discuss characteristic based models (as class of models) for equity market returns  in  \cite{HB1996}.
 In our analysis we implement   the form which they propose:
\begin{eqnarray}
R_{i,t} = \alpha_t^C + \sum_k \delta_{k,t} \theta_{i,k,t-1} + \epsilon_{i,t},\label{eqn:haugen}
\end{eqnarray}
where $\theta_{i,t-1}$ is the $i$-th observable stock characteristic at time $t-1$, such as book-to-price, and the payoffs to the $k$-th characteristic at time $t$ is $\delta_{k,t}$.
The expected return at time $t-1$ for time $t$ is:
\begin{eqnarray}
\mathrm{E}_{t-1}[R_{i,t}] = \mathrm{E}_{t-1}[\alpha_t^C] + \sum_k \mathrm{E}_{t-1}[\delta_{k,t}] \theta_{i,k,t-1} + \epsilon_{i,t}.\label{eqn:exphaugen}
\end{eqnarray}

For the comparison, the characteristic factors incorporated were size (market value) and value (book-to-price) attributes.

We note that the forms used in  Eqn. (\ref{eqn:risk}) -  Eqn.
(\ref{eqn:exphaugen}) should not be confused with general form
considered in the Daniel and Titman \cite{DT1997,DTW2001}, where
the risk model is purely explanatory and has {\em no time-lags}:
\begin{eqnarray} \mathrm{E}_{t-1}[R_{i,t}] = \alpha_t + \sum_k
\delta_{k,t} \theta_{i,k,t-1} + \sum_j \beta_{i,j,t-1}f_{j,t-1}.
\label{eqn:DT} \end{eqnarray} Furthermore, the formulation of Eqn.
(\ref{eqn:DT}) requires additional regressions to estimate
expectations of both prior characteristic and risk factor
loadings.

The forms used in  Eqn. (\ref{eqn:risk}) -  Eqn.
(\ref{eqn:exphaugen}) are consistent with our investigation to
test the ability of stock characteristics to act as predictors in
comparison with analogous risk factors. In particular,  we
consider the ability of the factor mimicking portfolios to predict
future returns (via dynamically changing loadings) rather than the
ability of the loadings themselves to act as predictors
(independent of the factor-mimicking portfolios). This latter
formulation has been explored elsewhere \cite{G2007}.

These equations are well suited for multi-factor regression
analysis with out-of-sample testing using small frequently
reformed data sets for rolling time-frames:  In order to compare
the risk-return profiles of CBM models described in  Eqn.
(\ref{eqn:haugen}) with corresponding risk factor models described
in Eqn. (\ref{eqn:risk}), ex-ante predictions from Eqn.
(\ref{eqn:exprisk})  and Eqn.  (\ref{eqn:exphaugen}) were used to
evaluate the risk-return profiles of the two models out-of-sample
. The expectations: $\mathrm{E}[\alpha_i^F]$,
$\mathrm{E}[\alpha^C]$, $\mathrm{E}[\beta_{i,j}]$ and
$\mathrm{E}[\delta_k]$ were assumed to be slowly varying functions
of time. This allowed us to estimate  $\mathrm{E}[\alpha_i^F]$ and
$\mathrm{E}[\beta_{i,j}]$ using time-series analysis and
$\mathrm{E}[\alpha_t^C]$ and $\mathrm{E}[\delta_{k,t}]$ using
cross-sectional analysis.

\subsubsection{The Fama and French APT model}

 We review  the construction of the three Fama-and-French risk factors
 (FF) which are proxied by {\em factor mimicking portfolios }
 based on  {\em book-to-price}  and {\em size} information. First, six
portfolios are constructed via  the intersection of three
book-to-price categories ($H\equiv$ High, $M\equiv$ Medium and
$L\equiv$ Low)  with two  {\em  size} categories ($B\equiv$ Big
and $S\equiv$ Small). These portfolios are designated $HS, \ MS, \
LS, \ HB, \ MB$, and $ LB$ \cite{FF1992,FF1995,FF1996,DT1997}.
These portfolios  are then used to capture the disentangled {\em
size} and {\em value} effects as factor mimicking portfolios:

\begin{eqnarray}
R_{_{SMB}} &=& \frac{1}{3} \left( {(R_{_{HS}} + R_{_{MS}} + R_{_{LS}}) - (R_{_{HB}} + R_{_{MB}} + R_{_{LB}})} \right), \\ \label{eqn:SMB}
R_{_{HML}} &=& \frac{1}{2} \left( {(R_{_{HB}} + R_{_{HS}}) - (R_{_{LB}} + R_{_{LS}})} \right) \label{eqn:HML}.
\end{eqnarray}

Here the risk factor variation $SMB$ is the excess return for small relative to large capitalized stocks (corrected for value), while the variation of $HML$ is the excess return of high value stocks relative to low value stocks (adjusted for size-effect contributions).

The resulting FF risk factor model for the $i$-th stock takes on the usual explanatory form
\begin{eqnarray}
& & R_{i,t} - R_{_{rfr},t} \\ &=& \alpha_i +
\beta_{i,_{Mkt}}(R_{_{Mkt},t} - R_{_{rfr},t}) + \beta_{i,_{HML}}
R_{_{HML},t} + \beta_{i,_{SMB}} R_{_{SMB},t} + \epsilon_{i,t},
\nonumber \label{eqn:FF}
\end{eqnarray}
where the risk-free rate of return is denoted by $R_{rfr,t}$ at time $t$, the bias is $\alpha$ and the factor loadings are given by the respective $\beta$'s. This is not the same form as the return-prediction mode for APT risk factors  used in Eqn. \ref{eqn:exprisk}. It is the predictive formulation that is investigated here. We note further that  it is the predictive power of the characteristics relative to the factor loadings that is explicitly tested for in  Daniel and Titman \cite{DT1997,DTW2001}, rather than the predictive power of the factor mimicking portfolio's as loaded by the $\beta$'s.

\subsection{Model implications: Arbitrage and linear vs nonlinear pricing kernels}




From a pragmatic point of view it may be that characteristic based
models make better predictions, given the vagaries and noise of
financial markets and financial data, simply because they are
better representations of available measurements.

Calibration to APT risk-factors is rewarded with linearity of the
pricing kernel.  A {\em linear pricing kernel } yields the price
of an asset as a scalar product of
 a representation of future payoffs (including contingent claims)
 with numbers which calibrate how the payoffs impact current asset prices.
 It's equivalence to the reasonable assumptions of NA pricing follows
by translating the NA into concise mathematical assumptions.

In particular, by combining NA with some simplifying assumptions
on the nature of investment returns in a finite (and complete)
state-space, an extremely elegant pricing framework can be derived
from a foundational theorem of early 20$^{th}$ century Hilbert
space theory \cite{Riesz1909}. The Riesz Representation theorem
provides the mathematical foundation for the following key
equivalences in the Fundamental Theorem of Asset Pricing and the
Pricing Representation theorems \cite{CR1976, DR2003}:
\begin{enumerate}
\item the market does not admit arbitrage opportunities
\item there exists a positive linear pricing rule for the relationship between
asset prices and future state-dependent payoffs ,
\item there exists a pricing measure under which risk-neutral discounted securities prices are
martingales
\item there exists of a positive pricing  kernel which provides the connection between positive probabilities of
events occurring and the payoffs of those events.
 \end{enumerate}

A key benefit of this approach is that  it facilitates a relatively simplified methodology to
pricing derivative claims which have nonlinear payoffs.

No-arbitrage has been generalised to more realistic continuous
(infinite-dimensional) and incomplete market settings
\cite{HK1979,HP1981} to the assumption of
no-free-lunch-with-vanishing-risk (NFLVR) \cite{DS1994}. The
latter can be interpreted as the condition whereby
 it is impossible to devise a potentially profitable {\em trading
 strategy}
which never  loses money.

While the failure of NA or NFLVR negates the existence of any
reliable market pricing kernel, the validity of a CBM does not
necessarily imply the existence of any arbitrage opportunities. In
fact it follows from arguments similar to those used in
\cite{DR2003}, that  it is also possible to construct a linear
pricing kernel in the case of a finite dimensional market which
admits a zero-alpha CBM.

In the literature,  stock specific characteristics  have been
interpreted to be {\em non-risk } determinants of asset prices
\cite{BCS1998}. Our view is that the effectiveness of these
variables points to a reality of {\em non-linear pricing
kernels}\footnote{ Giving up linearity of the pricing kernel poses
a further challenges for finance theory. In particular, if it is
possible to construct a NA characteristic based model which offers
more comprehensive and consistent market modelling, then the
result that firms are independent of their financial structure
\cite{MM1958,MM1961} may require reinterpretation.}. For example,
Daniel and Titman find $\alpha \neq 0 $ for both FF and CBM type
models investigated \cite{DT1997}. Such results imply that
$\alpha$ must be a nonlinear function of risk factors for a risk
pricing perspective to be preserved.

Departures from linearity may also be interpreted as manifesting
themselves in the $\beta$ of a factor-pricing model, where for example, the
$\beta$'s may be  functions of the characteristics that underpin the
factors. In this case, the model can remain linear in the
factors, but becomes non-linear in the characteristics in an
auto-regressive manner via factor coefficients \cite{FH1999}.

The notion of a nonlinear pricing kernel is consistent with the more general reflection of information in prices.
Motivated by the same considerations as Ross, Fama and French in their development of the linear APT risk factor approach, Bansal and Viswanathan (1993) and Harvey and Kirby (1995) investigated the theory of nonlinear pricing kernels in \cite{BV1994, HK1995}. Ferson and Harvey showed how to   calibrate the nonlinear (conditional) APT approach in \cite{FH1999}, while Chernov developed a  more general nonparametric calibration of nonlinear kernel models in \cite{Chernov2003}.




From an investors perspective, decisions are rewarded partly as
compensation for taking on the risk, which can be proxied by
appropriately specified factors,  as well as for removing possible
mispricings (arbitrage opportunities) identified by their
investment model.

We note that behavioural economists have not been convinced by the
risk-based explanation of the FF model \cite{LSV1994} and have
suggested, for example, that behavioural biases in the forward
estimation of earnings could explain the value premium found by
Fama and French \cite{FF1992}.  Similar criticism could be
directed to characteristic based models.

Thus, as asset managers \textit{chase alpha} by following almost identical strategies to identify possible
mispricings, the collective impact of their
investment strategies in the same assets can impact price formation.
This would be consistent with the extensive literature emanating
from theorists on the SA market, who have promoted characteristic approaches (\cite{vRR2003a, vRR2003b, vRR2004} and subsequent investigations in that journal).

We also note that the size effect can explained as a portfolio
level phenomena by following arguments given in the stochastic
portfolio theory (SPT) of Fernholz \cite{Fernholz1998, FK2006}.
While the SPT approach does not offer insight into the existence
or nature of pricing kernels, it does expose how capital entering and leaving the market impacts price as a function of cross-border cash-flow risk \cite{WG2007}.




In the next sections we present a period in the SA market for
which characteristic models have offered higher returns for lower
volatility, with the possible interpretation that APT models were
not able to price in as much information available to investors as
CBM.

\section{Data, market context and related investigations}

Our investigation is based on monthly data for stocks listed on the Johannesburg Stock Exchange
(JSE) between 1994 and 2007.

\subsection{Data and market context}

Our data set runs from 31 Jan 1994 to 30 April 2007, complementing
previous studies and including the bull-market post 2003, a period
in which  the market factor made a considerable contribution, as
well as  the peak in a value cycle in mid-2005.

We consider three distinct universes of stocks by using market
value to select the largest 50, 100 and 250 stocks respectively,
each month. The Top 250 stock universe was the largest universe
which could be kept sufficiently constant in stock number for the
study. We regarded this as  the simplest approach to creating a
representatuve universe of JSE as a whole for the out-of-sample
comparison of the APT and CBM models\footnote{Our data treatment
avoids the use of liquidity screens,  used in preceding analysis
of the same \cite{G2007, WG2004, WG2008}}.   In this paper we
report primarily on the Top 250 stock universe since the results
are similar across all universes including, yet surprisingly, the
relatively small Top 50 universe.

Continuous-time model returns (log-returns) were used. The proxy
for the cash asset was the current 3 month NCD (negotiable cash
deposit). This is the closest SA equivalent to US 3-month treasury
bills. South African NCD's are quoted at discount yield (NACQ).
This required that we construct a monthly price index using the
publicly quoted yield time-series for the instruments by computing
the face value and re-balancing monthly.

 Zero-valued prices and returns  were treated as missing
 data in all the sorts and regressions so that the lack
of meaningful numbers would not bias the analysis for  other
stocks and variables where data existed. In particular,  we did
not use interpolation methods, nor did we systematically exclude
data for some stocks due to missing data for others. Infinite
characteristic values were also treated as missing data in all the
sorts and regressions.

The data preprocessing follows Haugen and Baker \cite{HB1996} by
first winsorising the data at 3 standard deviations and then
z-scoring the resulting data. Missing data is excluded at the
level of the algorithms employed, and as such we have avoided
excluding entire data rows if missing data is encountered because
of data sparsity (rather than  the occurrence of non-trading
days).

 A synthetic market weighted index was computed from market values.
The CAPM $\beta$'s were computed relative the synthetic market
index. Such a market index was computed for each of the three
universes and re-balanced monthly. This was useful in that it
ensured that the market index was defined in terms of the same
universe of stocks as the study itself as well as ensuring
meaningful comparison within the 3 stock universes.

We restricted ourselves to a small set of input variables for the
characteristic models: a total return index, historic dividend
yield, price-to-book, volume traded, market value and earnings
yield. The data was sourced from Thomson-DataStream. Although a
more extensive list of characteristics was available  and
commercial applications of such CBM models include forward broker
information\footnote{A typical stable commercial characteristic
based model for the South African market could include most of the
following factors (perhaps less if volatility becomes a concern):
total return, log-size, price, book-to-price, cash-to-price,
dividend yield, earnings yield, 1-year forward earnings yield,
2-year forward earnings yield, 1-year forward earnings growth,
earnings torpedo (change from latest earnings to next consensus
earnings), neglect (negative log of number of analysts covering a
stock), earnings revision, earnings downgrade, earnings upgrade,
broker recommendation (buy, hold, sell), low price, payout ratio,
3-month momentum, 6-month momentum, 9- month momentum, 1-month
momentum (smoothed), currency plays (dummy for USD/ZAR
exposure).}, the intention here is to focus on value and size in
the context of  $HML$ and $SMB$ constructions for the Johannesburg
Stock Exchange.

Two momentum characteristics were also computed using the total
return index. Specifically,  a long-term 12-month momentum, lagged
1 month prior to formation date, and a short-term 3-month momentum
signal, also lagged 1 month prior to formation date, were
computed. The limited length of the data sets and the relative
sparsity of the sort data made momentum signals longer than 2
years impractical.

A unique problem with the South African market is the dominant
effect that dual listed stocks have. Dual listed stocks whose
primary listings were not on the Johannesburg Stock Exchange where
included and the book-to-price and earnings yield of the primary
listed entities where used in the analysis \footnote{This included
dual-listed counters such as Anglo American, BHP-Billiton,
Richmont, SAB-Miller, Old Mutual, Liberty Life and Investec. The
list was constructed using the JSE listed ISIN codes to find the
primary listing exchange and tickers. With this information the
correct primary listing characteristics could be sourced. By using
price ratios in the primary listing's currency  one avoids
additional uncertainties arising when converting the
characteristics, such as earnings and book-value, to local
currencies.}.

\subsection{Related investigations}
There have been several studies relating to characteristic based
models on JSE listed stocks since \cite{FP2000, vRR2003a,
vRR2004}. More recently, we considered the impact of foreign
portfolio investment, as a risk factor for emerging markets
\cite{WG2007}, via the stochastic portfolio theory factorisation
of Fernholz \cite{Fernholz1998}.

In  \cite{FP2000, vRR2003a, vRR2004} the authors argued that the
cross-section of returns on the JSE is explained by
characteristics (attributes) rather than risk factors for
correlated price-to-earnings and size risk factors. Our approach
is different however, since  \cite{vRR2004} constructed the
following: a portfolio $SLL$ or {\em small-less-large}, driven by
the difference  between small and large cap returns,  as a risk
factor capturing the size effect, and a portfolio {\em LLH} or
{\em  low-less-high}, driven by the difference between low and
high value stock returns. Their factors are uncorrected for
possible correlation, which is a key ingredient in the FF risk
factor constructions \cite{FF1996, DT1997}.

Daniel and Titman  recommend the use multi-factor regressions on
triple sorted portfolios\footnote{The use of multi-factor
regressions in the context of triple sorted portfolio intercept
tests was  also presented in \cite{G2007}.} as a step for dealing
with the errors-in-variables problem which portfolio sort based
testing procedures have \cite{DTW2001}.  Understanding  of the
method is important because it provides a key motivation for our
review of the FF model for South African data. We note that
\cite{vRR2004} used single factor regressions in the context of
double sorted portfolios. However, the theoretical context of
their contribution was not that of discriminating between risk and
characteristic-based models but rather that of identifying a more
appropriate form of asset pricing model specification  based on
characteristics (attributes).

The sample studied by Fraser and Page \cite{FP2000} was from 1973
through 1997 and examined financial and industrial stocks with a
focus on price-to-earnings, dividend yield and market value. They
argued that price-to-earnings was the key determinant of the
cross-sectional price differences at that time. The sample studied
by \cite{vRR2003a} ran 1990 through 2000 and also focused on
price-to-earnings and market value. They considered  a more
extensive set of characteristics with commercial applications in
fund management in view. This provided one of the first screens of
characteristics and aligned with their stated objective of finding
a more appropriate asset pricing model for the JSE.


In our investigation,  we recover an apparent {\em value effect}
in three different stock universes, namely the Top 50, Top 100 and
Top 250 stocks, as reformed monthly. The {\em size effect} was
more prominent in  the Top 50  and Top 250 stock universes. Prior
work on the size effect on US common stocks \cite{Banz1981} was
extended to the JSE by Page and Palmer \cite{PP1993}. Similarly,
the effect of price-to-earnings \cite{Basu1983} and price-to-book
\cite{Stattman1980} has been extended to the JSE
\cite{FP2000,vRR2003a}. Various ad hoc investigations on price
premia in the same market are also documented in the literature.

We add to this discussion in the context  that the South African
market has hierarchical aspects: ALSI 40 (largest 40 stocks)
attracts international investors while simultaneously being
important to domestic investors, who also focus on the ALSI
(largest 165 stocks). We promote a nonlinear pricing kernel
perspective to reconcile the findings in favour of direct
characteristic based pricing with no-arbitrage pricing paradigms.


\section{Results and discussion}


\subsection{Risk and return comparisons between model types}

We compare a 3-factor APT model, using our FF factors, and  a CAPM
model with two characteristic based models (CBM), one without
momentum (CBM \#2 in Table \ref{tab:payoffs}) and one with
momentum (CBM \#6 in Table \ref{tab:payoffs}).

For Figure \ref{fig1:riskreturn}, returns for each model were
sorted\footnote{The sorts where carried out using a symmetric
quantiling algorithm that kept the sort symmetric around the
middle quantile for odd numbers of quantiles, and symmetric around
the two middle quantiles for even numbers of quantiles. The
algorithm was also tailored to ignore missing data and cope with
listings, delistings  and illiquid instruments that may have had
no trade or fundamental data available on a given month. This
quantiling algorithm was then used to sort stocks into quintiles.}
into five quintiles, with highest expected returns binned into the
$1^{st}$ quintile. Best fit lines for corresponding realised
returns and realised volatilities were then plotted.

  The CAPM and FF risk models yielded ex-post returns
which were less than those from two CBM, but at higher
volatilities. This is evident in Figure \ref{fig1:riskreturn} for
the Top 250 universe, but holds for all three stock universes
studied and demonstrates that  one could generate higher returns
with lower risk by using CBM and selecting stocks whose returns
fall into the $1^{st}$ quintile. A more rigorous statistical
treatment using the Daniel and Titman \cite{DT1997, DTW2001}
approach was given in \cite{G2007} and is consistent with the
visual results shown here.

 Risks, measured by volatility, vary little across the quintiles
for CAPM and FF models, but are significantly higher in the
$4^{th}$ and $5^{th}$ quintiles for their characteristic based
counterparts. This result is robust across the three different
universes of stocks. We note further that loadings to the market
factors in the APT and CAPM models were higher than the
corresponding market factor loadings in the CBM.


Hence, we claim that CBM have been  effective even in fairly small
concentrated data sets, such as the ALSI 40 universe, as well as
being effective in the larger universes. This demonstrates that
investors have not necessarily rewarded for risk in the sense of
APT or CAPM on the JSE.  These results holds across the Top 250,
Top 100 and Top 50 stock universes, providing a key finding for
JSE for the pre-global crisis period considered and corroborating
the findings of \cite{Haugen1998}.

In \cite{G2007}, which implemented the triple-sort approach of
Daniel and Titman \cite{DT1997} to compare the models, it was
found that generally $\alpha \neq 0$ for the  FF models. Assuming
that the FF factors provided a robust model, this points to
possibility that the stock $\alpha$'s were nonlinear functions of
risk \cite{FH1999}.

\subsection{JSE results for the Fama and French risk factor model}

\begin{figure}
\includegraphics[width=12cm]{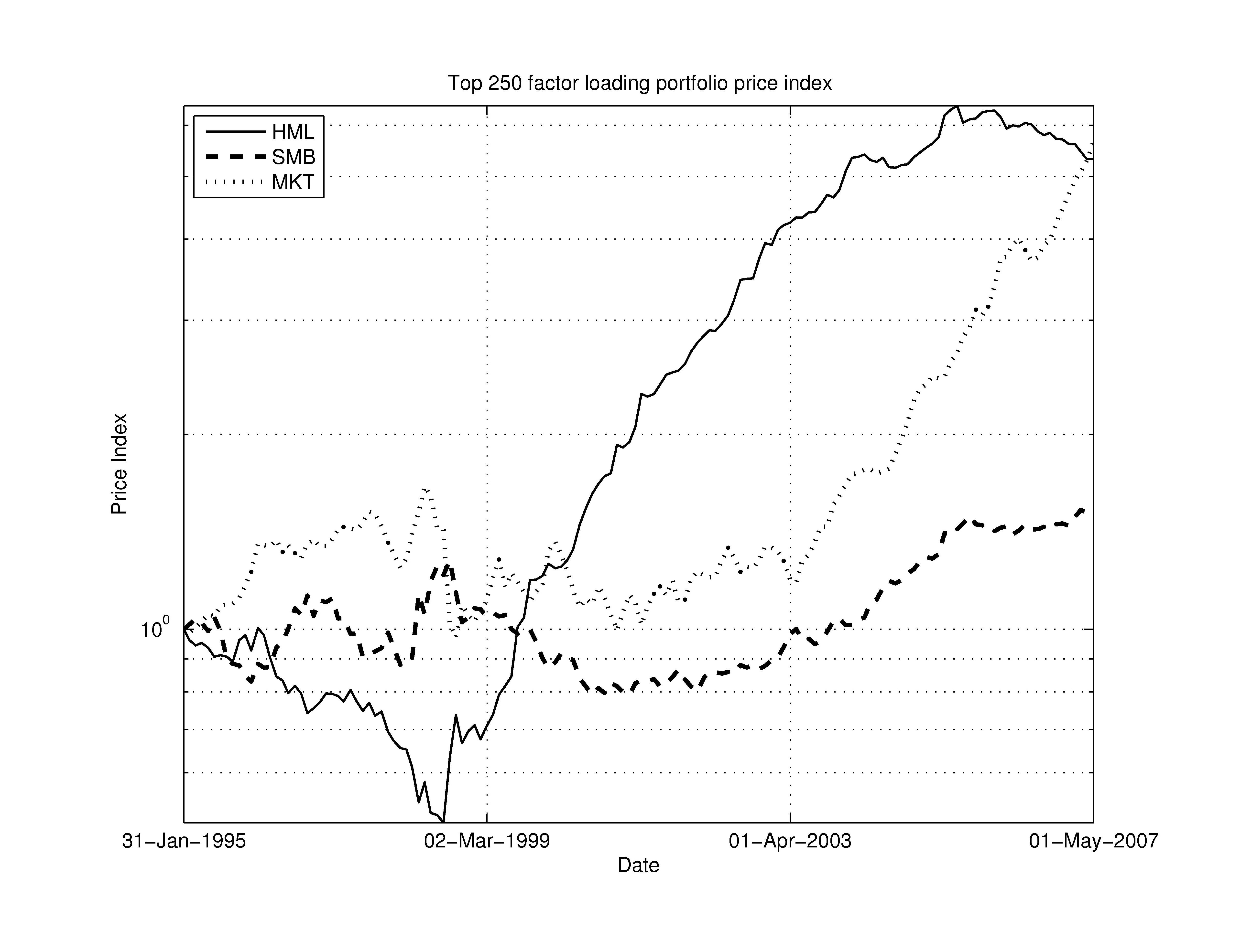}
\caption{\small The Top 250 {\em HML} (High-Minus-Low)and {\em
SMB} (Small-Minus-Big) factor mimicking portfolios formed from the
6 intersection portfolios as in Eqn's. 5 and Eqn. 6 and reformed
monthly using only the largest 250 stock each month out of 472
stocks listed on the JSE between 31 January 1994 and 30 April 2007
are shown. The {\em HML} (value) factor mimicking portfolio is
given by the solid line, the {\em SMB} (size) factor mimicking
portfolio is given by the dashed line and the market portfolio
{\em Mkt} is given by the dotted line.}\label{fig1:HML_SMB}
\end{figure}

The cumulative factor returns for the { value} ({\em HML}), size
({\em SMB}) and market (Mkt)  factors are given in Figure
\ref{fig1:HML_SMB} for the Top 250 universe of stocks. The market
factor return was constructed using stock  market capitalisation
values and the factor was re-balanced monthly. The fundamental
factor (market cap data) dates were shifted 1 quarter backward in
time relative to the date the information is documented in the raw
data, since the latter is done with a 1 quarter time lag.
The $\beta$'s were estimated in 6 year rolling windows (72
months).

The study of the Top 250 stock universe uncovered a delayed size
effect. Assessment of  the {\em SMB} factor returns \cite{G2007}
showed a peak 12 months after portfolio formation. By inspection,
there was limited return advantage apparent in the time-series
behaviour of the {\em SMB} factor shown in Figure 1. Nevertheless,
there was still a discernible size-effect.  The value effect is
clearly evident from 1998 through to mid 2005. The importance of
the market factor after 2003 is  visually apparent from the
nominal performance of the factor mimicking portfolio in Figure 1.

\subsection{Observations for characteristic based models}

Results are summarised for 14 characteristic based models in Table
1 for the 250 stock universe studied. Table entries are the median
characteristic payoffs .

As expected, we find a broad positive loading to book-to-price and
a negative loading to size. This is in agreement with the analysis
using the {\em HML} and {\em SMB} factor loading portfolios. The
value effect increases with increase in universe size, while the
size effect diminishes. This is surprising, since  one would
expect the size effect to be more meaningful in a larger universe
of stocks. When controlling for dividend yield the size effect is
diminished in the Top 50 universe but not significantly diminished
within the Top 100 and Top 250 universes of stocks. The size
effect is also diminished in the presence of momentum.

\normalsize
\begin{table}
\begin{footnotesize}
\begin{tabular}{cccccccccc}
\hline
& \multicolumn{9}{c}{Median Payoffs to Z-scored characteristics [units of\%]}\\
\hline
Model \# & $\alpha$ & $\delta_{_{BVTP}}$ & $\delta_{_{MV}}$ & $\delta_{_{Mkt}}$ & $\delta_{_{MOML}}$ & $\delta_{_{MOMS}}$ & $\delta_{_{EY}}$ & $\delta_{_{DY}}$ & $\delta_{_{VOL}}$ \\
\hline
1 & 1.93 & 0.39 & -0.15 & & & & & & \\
2 & 1.92 & 0.41 & -0.17 & 0.01 & & & & & \\
3 & 1.96 & 0.57 & -0.15 &  & 0.36 & & & & \\
4 & 1.95 & 0.51 & -0.18 & 0.30 & -0.00 & & & & \\
5 & 1.96 & 0.47 & -0.13 & 0.74 & -0.51 & & & & \\
6 & 1.96 & 0.46 & -0.19 & 0.65 & -0.49 & -0.14 & & & \\
7 & 1.99 & 0.49 & -0.15 & 0.57 & -0.54 &  0.12 & & & -0.11 \\
8 & 1.94 & 0.41 & -0.25 & 0.10 & & & & & 0.04\\
9 & 1.75 & 0.44 & -0.06 & 0.60 & -0.52 & 0.43 & -0.19 & & \\
10 & 1.70 & 0.26 & -0.08 & & & & 0.16 & & \\
11 & 1.65 & 0.42 & -0.10 & & 0.49 & -0.52 & 0.26 & 0.13 & \\
12 & 1.85 & 0.36 & -0.20 & & & & & 0.04 & \\
13 & 1.86 & 0.50 & -0.07 & 0.49 & -0.47 & 0.24  & & 0.12 & \\
14 & 1.68 & 0.04 & -0.14 & 0.49 & -0.48 & 0.12 & 0.28 & 0.01 & \\
\hline
\end{tabular}
\end{footnotesize}
  \caption{\small Top 250 stock JSE characteristics based models. The characteristic factors are, from left
to right, the model bias, book-to-price, market value, market
factor, long term momentum, short term momentum, earnings yield,
dividend yield and volume traded. Model \#2 and \#6 are compared
with CAPM and the 3-factor Fama and French APT model in Figures 2
and 3. The dynamics of the payoffs for model \#14 (the medians are
presented here) is shown in Figure 2. }\label{tab:payoffs}
\end{table}
\normalsize

The book-to-price effect is correlated with the earnings yield
(inverse of price-to-earnings), and it is found that the
book-to-price effect can be reduced when controlling for earnings
yield in the presence of size. This apparent multi-collinearity
between book-to-price and earnings yield becomes less effective in
the presence of momentum. This broadly corroborates the identified
ability to substitute book-to-price with price-to-earnings and
size on the JSE \cite{vRR2003a}.

It is also noted the earnings yield and change in volumes traded
provide little additional explanatory power when in the presence
of price-to-book, market value, momentum and a market factor.
Hence, there appears to have been little additional advantage in
using volumes traded, given its  high multi-collinearity with
other factors. The market factor can be substituted with volume
traded.

We do find subtle dependencies on the dividend yield: when
controlling for the loading on the market portfolio one finds that
the effect of the dividend yield changes sign in the Top 50
universe. Long-term momentum is generally positive except when
controlling for the market portfolio. Short term momentum is
negative except when controlling for dividends, earnings and
volumes.

\begin{figure}
\includegraphics[width=12cm]{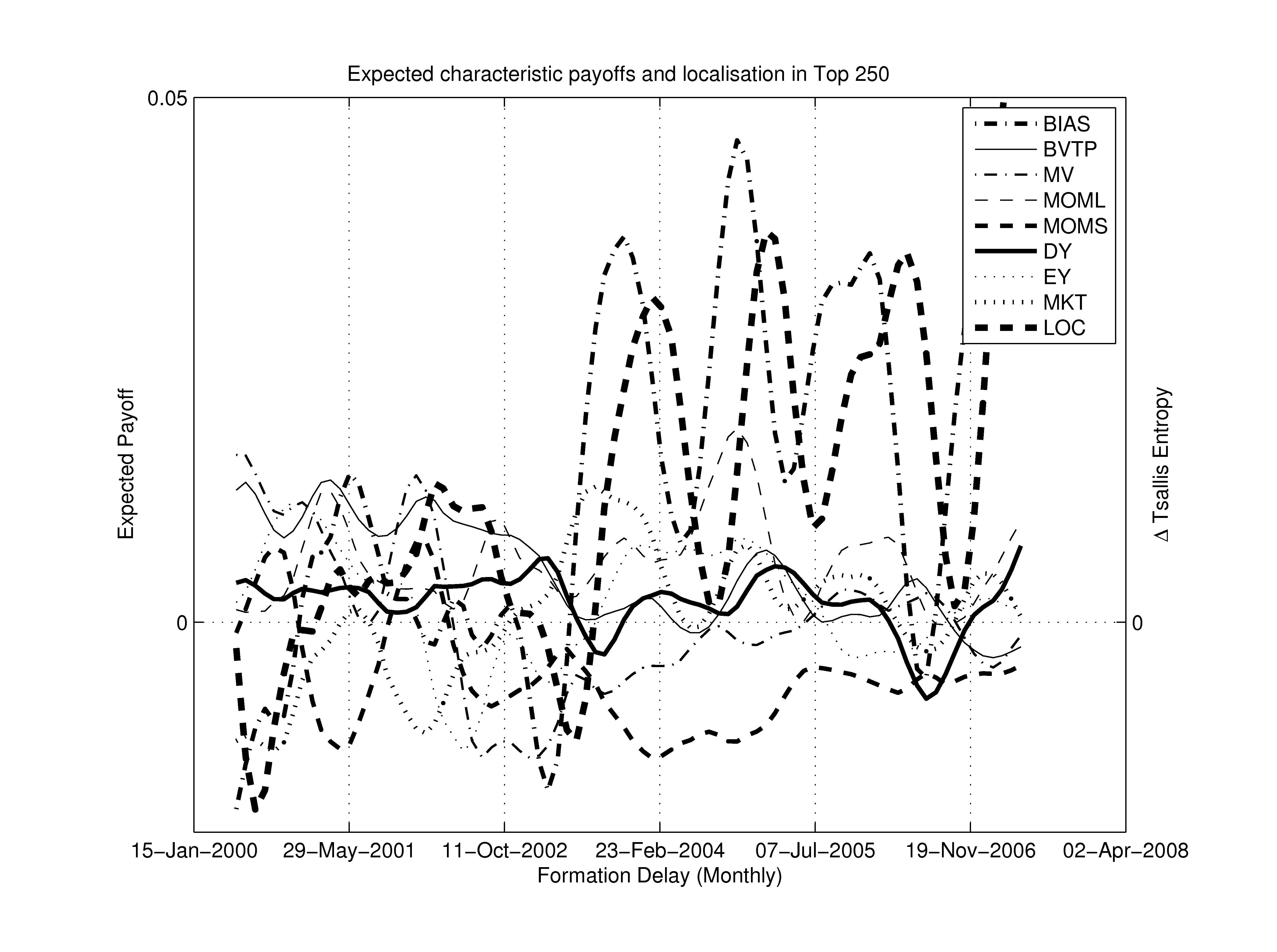}
\caption{\small The Top 250 dynamics of the smoothed payoffs to
the various characteristics used in CBM \#14 from Table
\ref{tab:payoffs} against the change in Tsallis entropy (LOC),
which is a measure of market wide localisation (the more localised
the market the fewer opportunities there are because portfolio
market values are more concentrated on fewer stocks). The dynamic
behaviour of the characteristic payoffs is apparent.
}\label{fig1:payoff_dynamics}
\end{figure}
\begin{figure}
\includegraphics[width=12cm, height=5.5cm]{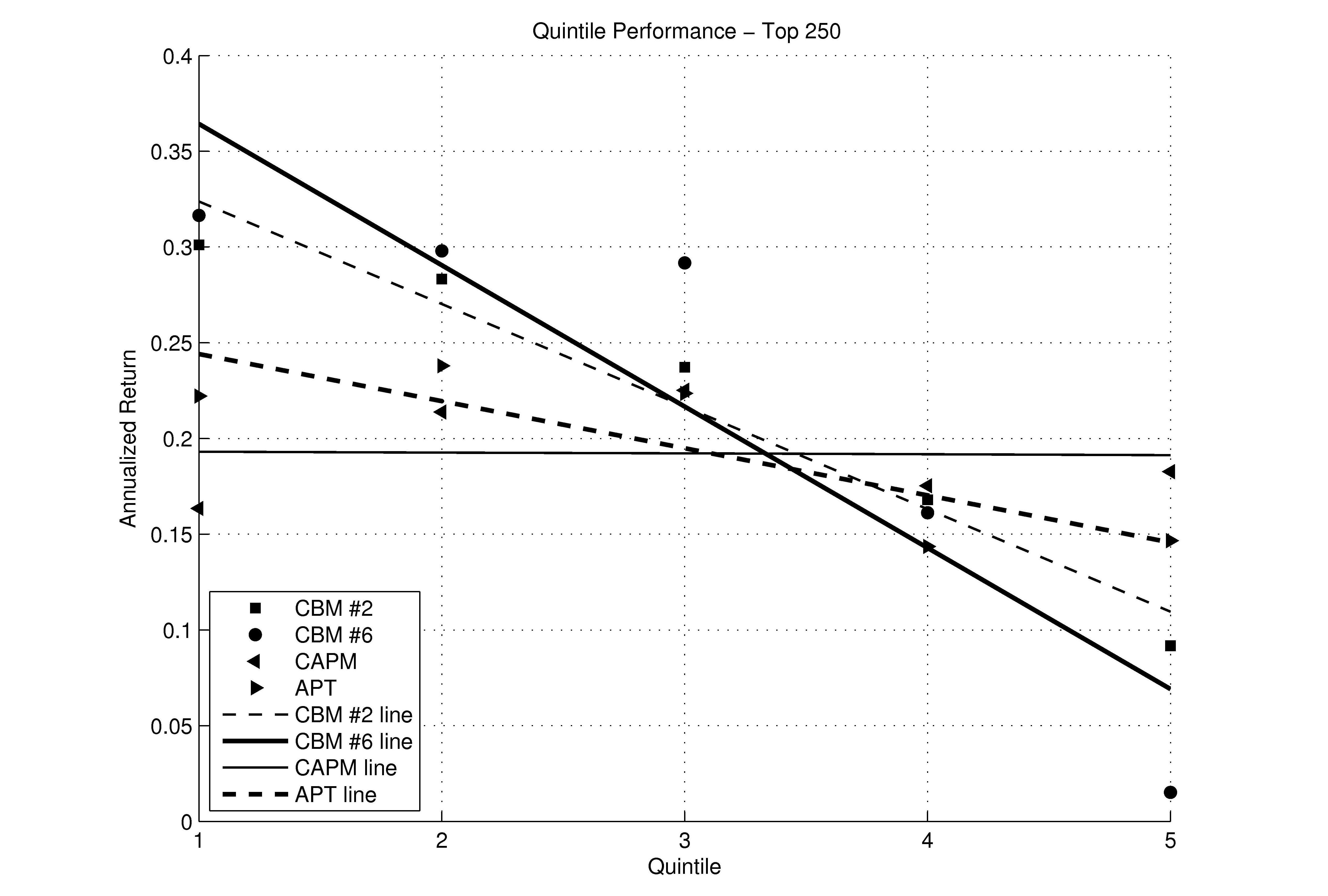}
\includegraphics[width=12cm, height=5.5cm]{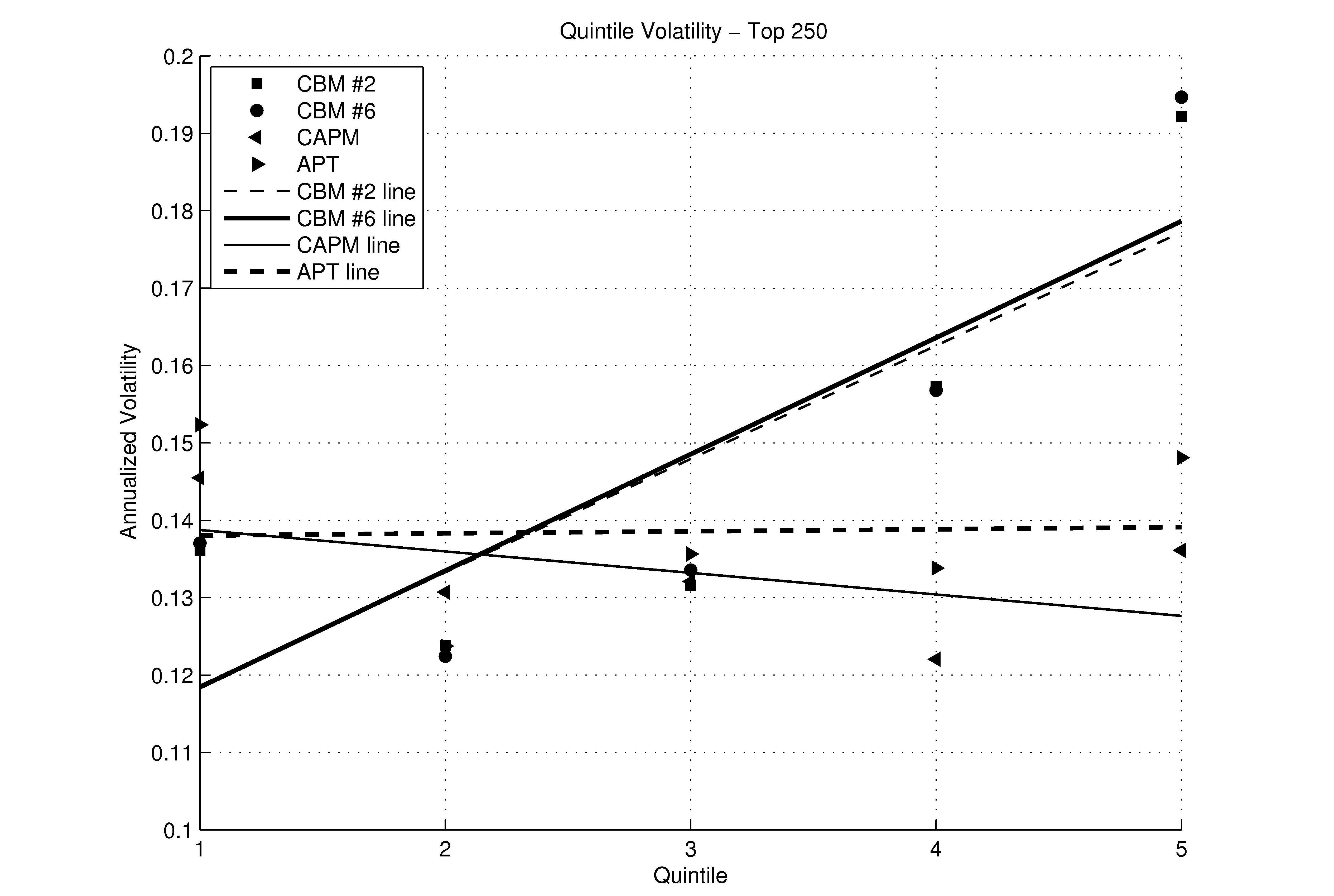}
\caption{\small Stocks were quintiled on expected return, with the
highest in the first quintile. The upper graph has the average
annualised returns in the top 250 stock universe for each of the 5
quintiled portfolio's of two characteristic based models and two
risk based models is given. The lower graph depicts the average
volatilities. The first characteristic based model (CBM) uses
market capitalisation, book-to-price and a market factor and
corresponds to CBM \#2 in the Table 1 and is denoted by solid
squares with a thin dashed best fit line. The second CBM is model
\#6 from Table 1, it includes momentum signals and is denoted by
dark circles with a solid line running from the upper left to
lower right. The risk based models (RBM) are the 3-Factor APT
model using the HML, SMB and Mkt factors and CAPM, denoted by
right triangles and a solid dash best fit line and left triangles
and the solid horizontal best fit line respectively. The CBM
models provide realised returns commensurate with the quintile
sorts from highest expected return quintile 1 through to the
lowest expected return quintile 5. As such the CBM model are good
predictors of portfolio wide return commonalities, the RBM models
are not. The lower graph has the volatilities of two risk based
asset pricing models and two characteristic based pricing models
in the top 250 stock universe. The CBM models provide realised
volatilities inversely related to the quintile sorts from highest
expected return quintile 1 through to the lowest expected return
quintile 5. As such the CBM models provide lower volatilities in
quintiles with higher expected returns this is contrary to the RBM
models which expect higher returns with relatively higher
volatilities. }\label{fig1:riskreturn}
\end{figure}

The payoffs to the various characteristics have time dependent
oscillatory dynamics. This is graphically demonstrated in Figure
\ref{fig1:payoff_dynamics} for CBM model \#14 (whose median
payoffs values are in Figure \ref{tab:payoffs} for the full
sample). For example, prior to December 2001 there was a positive
payoff to market value.

This shifted to a negative payoff to market value a year later,
which can be attributed  to the December 2001 crash in the USD/ZAR
exchange rate. Similar pathologies can be seen for almost all the
characteristic payoffs,  demonstrating that price anomalies
associated with payoffs to characteristics were reasonably
short-term.

On a practical note, using an extensive list of characteristics
can lead to data-mining effects. This is particularly important in
a concentrated market such as the JSE. This may arises when
multi-collinearity swamps the model estimation process with noise
effects. It is our view that a model with more than about 10
factors in the SA market will be prone to this flaw.  It is also easy to
select a particular window of data for which the model happens to
work well. This type of error is difficult avoid in a bull market
as almost all the characteristics become coupled to momentum.


We recommend that practitioners opting to use CBM
carry-out tests for randomness in the model outputs in addition to
the usual out-of-sample back-testing. We also recommend scenario
analysis with characteristic shocks to further understand dynamics
under noisy conditions due to the presence of oscillatory cycles
(see Figure \ref{fig1:payoff_dynamics}). 

A related concern is that
associated with the liquidity premium. Characteristic based models
constructed with many too factors can result in the identification
of a liquidity premium which cannot be traded. 
Finally, we note that equity premia have been shown to include exogenous factors such as currency flows during regional and global crises \cite{WG2007}.

\section{Conclusion}

We have demonstrated that cross-sectional characteristic models have
yielded portfolios with higher excess monthly returns but lower
risk than their risk based factor model counterparts on the JSE between 1994 and 2007.  The outcome is consistent with the Daniel-and-Titman triple sort comparison carried out in \cite{G2007}.
Thus, investors appear to have been rewarded for the self-consistent use of information as
proxied by stock characteristics.

This may  simply have been a reflection of the way information and
risk were priced. In particular, CBM may have priced information
more efficiently if stock specific accounting variables were used
by many investors in that period.  With such models popularised in
academic literature, the price impact of investors using similar
strategies may have influenced price fluctuations in traded
assets.

We have focussed our attention on detecting how size and value
information were reflected in stock prices and of the two models
investigated, the cross-sectional characteristic model was the
better predictor of returns. Extending the pricing kernel further
to incorporate  more prevailing market information, including
other stock characteristics, exchange-traded derivative prices and
credit, liquidity and other risks in an incomplete market, would
more likely confirm a nonlinear price transition kernel.

Characteristic based models are typically nonlinear, but can still
be consistent with assumptions of NFLVR for a market with
non-unique risk-neutral pricing solutions \cite{Chernov2003,
HK1995}.  At the same time, evidence of CBM information pricing
does not imply that a market is free of arbitrage.

In the context of an aggregate influx of capital into the JSE
\cite{WG2007} for the period investigated, this paper highlights
that it was possible to select outperforming portfolios by
incorporating information based on  value and size criteria
directly, as compared to pricing risk information via
Fama-and-French time-series factors. This contributes to the
understanding of how information and risk are priced within a more
general Markovian perspective of  markets.

\section*{Acknowledgements}

The authors thank the participants of the 2008 South African
Finance Association conference, held that the Graduate School of
Business at the University of Cape Town, and participants of the
Summer School on Mathematical Finance held at the African
Institute for Mathematics in February 2008. Criticism from proponents
of strictly linear APT modelling served as motivation for us to reconcile the
evidence for characteristic based pricing.

This research was partially funded by a Carnegie Foundation research grant
administered by the University of the Witwatersrand and an National Research
Foundation NPYY Grant [Number 74223].

\begin{thebibliography}{00}

\bibitem{AG1994}
Allen, F., Gale, D.,  Financial Innovation and Risk Sharing, MIT
Press, 1994

\bibitem{Ball1978}
Ball, R., (1978),  Anomalies in Relationships Between Securities'
Yields and Yield-Surrogates, {Journal of Financial Economics}
6, 103-26.

\bibitem{BV1994} Bansal, R., Viswanatham, S., (1993)  No arbitrage and arbitrage pricing: A new approach, { Journal of Finance}, Vol. 48, No. 4, 1231-1262



\bibitem{Banz1981} Banz, R., (1981) The relationship between return and market value of common stock, Journal of Financial Economics, 9, 3-18

\bibitem{Basu1983} Basu, S., (1983) The relationship between earning yields, market value, and return for the NYSE common stocks: Further evidence, Journal of Financial Economics, 12, 129-156

\bibitem{BK2002} Barr, G., Kantor, B., (2002) The South African Economy and its Asset markets - an integrated approach, { South African Journal of Economics,} 70 (1), 53-78.

\bibitem{Black1972} Black, F., (1972) Capital Market Equilibrium with restricted borrowing, Journal of Business 45, 444-454

\bibitem{BCS1998}
Brennan, M.J., Chordia, T., Subrahmanyam, A., (1998) Alternative factor speci�cations, security characteristics,
and the cross-section of expected stock returns, { Journal of Financial Economics}, 49  345-373


\bibitem{Chernov2003} Chernov, M., (2003) Empirical Reverse Engineering of the
Pricing Kernel, { Journal of Econometrics,} 116, 329-364



\bibitem{CR1976} Cox, C., Ross, S. A., (1976) The valuation of options for alternative stochastic processes, Journal of Financial Economics, 3, 145-166.

\bibitem{DT1997} Daniel, K., Titman, S., (1997) Evidence on the Characterstics of Cross Sectional Variation in Stock Returns, Journal of Finance, 52, 1-33

\bibitem{DTW2001} Daniel, K., Titman, S., Wei, J., K., C., (2001) Explaining the cross-section of Stock Returns in Japan: Factors or Characteristics? Journal of Finance, 55, (2), 743-766

\bibitem{DS1994}
Delbaen, F., Schachermayer, W., (1994) A general version of the
fundamental theorem of asset pricing, Mathematische Annalen 300,
463� 520.


\bibitem{DR2003}
Dybvig, P.,  Ross, S.A.,  Arbitrage, State Prices and Portfolio Theory, in { Handbook of the Economics of Finance},  Constantinides, G.M., Harris, M., Stulz, R., Eds., Elsevier, 2003

\bibitem{FF1992} Fama, E., French, K., (1992) The cross-section of Expected Stock Returns, Journal of Finance, 47, 427-465

\bibitem{FF1993} Fama, E. F. and K.R. French (1993) Common risk factors in the returns on stocks and bonds, Journal of Financial Economics 33, 3-56.

\bibitem{FF1995} Fama, E., French, K., (1995) Size and book-to-market factors in earnings and returns, Journal of Finance, 50, 131-155

\bibitem{F1996} Fama, E., (1996) Multifactor Portfolio Efficiency and Multifactor Asset Pricing, The Journal of Financial and Quantitative Analysis, 31, 4, Dec. 1996, 441-465

\bibitem{FF1996} Fama, E., French, K., (1996) Multifactor explanations of asset pricing anomalies,  Journal of Finance, 51, 55-84

\bibitem{Fernholz1998} Fernholz, R., (1998) Crossovers, Dividends, and the Size Effect, Financial Analyst Journal, May/June 1998, 73-78

\bibitem{FK2006} Fernholz, R., Karatzas, I., (2006) The implied liquidity premium for equities, Annals of Finance, (2), 87-99.

\bibitem{FH1999} Ferson, W. E., Harvey, C. R., (1999) Conditioning Variables and the cross-section of Stock Returns, Journal of Finance, 56, 4, 1325-1360

\bibitem{FS2002} Firer, C., Staunton, M., (2002) 102 Years of South African financial market history,  Investment Analysts Journal, 56 (5), 2

\bibitem{FP2000} Fraser, E., Page, M., (2000) Value and Momentum strategies: Evidence from the Johannesburg Stock Exchange, Investment Analysts Journal, 51, 14-22



\bibitem{G2007} Gebbie, T., Wilcox, D., (2007) Evidence of characteristic cross-sectional pricing of stock returns in South Africa, Working paper (presented SAFA 2007, AIMS 2008)


\bibitem{Haugen1998} Haugen, R., (1998) Relative Predictive Power of the Ad Hoc Expected Return Factor Model and Asset Pricing Models, Applied Equity Valuation, Coggin T. D., Fabozzi, F. J. (Editors), John Wiley \& Sons

\bibitem{HB1996} Haugen, R. A., Baker, L. N., (1996) Commonality in the determinants of expected stocks returns, Journal of Financial Economics ,41, 401-439

\bibitem{HK1979} Harrison, J. M., Kreps, D. M., (1979) Martingales and arbitrage in multiperiod securities markets, Journal of Economic Theory, 20, 381-408

\bibitem{HP1981} Harrison, M., Pliska, S., (1981) Martingales and Stochastic integrals in the theory of continuous trading. Stochastic  Processes \& Applications, Vol. 11, pp. 215-260


\bibitem{HK1995} Harvey, C., Kirby, C., (1995) Analytic tests of factor pricing models, Unpublished note


\bibitem{JS2005} Jefferis, K., Smith, G., (2005) The changing efficiency of African Stock Markets, South African Journal of Economics, 73, (1), 54-67

\bibitem{LSV1994} Lakonishok, J., Schleifer, A., Vishny, R., W., (1994) Contrarian investment, extrapolation and risk, Journal of Finance, 49, 1541-1578

\bibitem{Lintner1965} Lintner, J., (1965) The valuation of risk assets and the selection of risk investments in stock portfolios an capital budgets, Review of Economics and Statistics, 47, 13-37


\bibitem{Matoti2009}
Matoti, L., Building a Statistical Linear Factor Model Gate and a
Global Minimum Variance Portfolio Using Estimated Covariance
Matrices, UCT MSc Dissertation, 2009




\bibitem{Merton1973} Merton,R.,C., (1973) An intertemporal capital asset pricing model, Econometrica, 41, 867-887

\bibitem{MM1958} Mogdigliani, F., Miller, H., M., (1958) The cost of capital, corporate finance and the theory of investment. American Economic Review, 48(3), 261-97

\bibitem{MM1961} Miller, H., M., Mogdigliani, F., (1961) Dividend policy, growth, and the valuation of shares, Journal of Business 34 (4),  411-433

\bibitem{NS2004} Nawalkha, S., K., Schwarz, C., (2004) The Progeny of CAPM,  Journal of Investment Management, 3, 2, 53-61

\bibitem{Nawalkha1997} Nawalkha, S., (1997) A multibeta representation theorem for linear asset pricing theories, Journal of Financial Economics, 46, (3), 357-381


\bibitem{PP1993} Page, M., Palmer, F. (1993) The relationship between excess returns, firm size and earnings on the Johannesburg Stock Exchange, South African Journal of Business Management, 22(3), 63-73



\bibitem{Riesz1909} Riesz, F., (1909) Sur les op$\acute{e}$rations fonctionelles lineaires, {  Comptes rendus de l'Acad$\acute{e}$mie des
sciences,} 149, 974 - 977


\bibitem{Ross1976} Ross, S., (1976) The arbitrage theory of capital asset pricing, Journal of Economic Theory, 13, 341-360

\bibitem{Ross2005} Ross, S.,  { Neoclassical Finance, } Princeton Lectures in Finance, Princeton Press, 2005

\bibitem{RR1994} Roll, R., Ross, S. A., (1994) On the cross-sectional relation between expected returns and betas,
Journal of Finance, 49, 101-121

\bibitem{Sharpe1964} Sharpe, W., F., (1964) Capital asset prices: a theory of market equilibrium under conditions of risk,
Journal of Finance, 19, 425-442

\bibitem{Stattman1980} Stattman, D., (1980) Book values and expected stock returns, The Chicago MBA: A journal of selected papers 4, 25-45

\bibitem{vRR2003a} van Rensburg, P., Robertson, M., (2003a) Style Characteristics and the cross-section of JSE returns, Investment Analysts Journal, 57, 7-16

\bibitem{vRR2003b} van Rensburg, P., Robertson, M., (2003b) Size, price-to-earnings and beta on the JSE Securities Exchange, Investment Analysts Journal, 58, 7-16

\bibitem{vRR2004} van Rensburg, P. , Robertson, M., (2004) Explaining the cross-section of returns in South Africa: attributes or factor loadings? Journal of Asset Management, vol. 4 no.5, 334-347

\bibitem{WG2004} Wilcox, D., Gebbie, T., (2004) On the analysis of cross-correlations in South African market data, Physica A, 344, 294-298


\bibitem{WG2007} Wilcox, D., Gebbie, T., (2007) Factorising Equity Returns in an Emerging Market Through Exogenous Shocks and Capital Flows,  SSRN: http://ssrn.com/abstract=2283486

\bibitem{WG2008}Wilcox, D., Gebbie, T., (2008) Periodicity and scaling of eigenmodes
in an emerging market, International Journal of Theoretical and
Applied Finance, 11, 7, 739-760



    \end{thebibliography}

\end{document}